\documentclass[apjl]{emulateapj}

\slugcomment{}

\shorttitle{The $\gamma$-ray emission region of 3C~111}
\shortauthors{Grandi}

\begin{document}


\title{The $\gamma$-ray emission region in the  FRII Radio Galaxy 3C~111}


\author{P. Grandi,  E. Torresi,}
\affil{Istituto Nazionale di Astrofisica -  IASFBO, Via Gobetti 101, I-40129, Bologna, Italy}
\and
\author{C. Stanghellini}
\affil{Istituto Nazionale di Astrofisica - IRA, Via Gobetti 101, I-40129, Bologna, Italy}

\begin{abstract}
The Broad Line Radio Galaxy 3C~111, characterized by a Fanaroff-Riley II (FRII) radio morphology,  is one of the sources of the  Misaligned Active Galactic Nuclei sample, consisting of  Radio Galaxies and Steep Spectrum Radio Quasars, recently detected  by the {\it Fermi}-Large Area Telescope.  

Our analysis of the  24-month  $\gamma$-ray light curve shows  that 3C~111  was only occasionally detected at high energies.
It  was bright at the end of 2008 and faint,  below the {\it Fermi}-Large Area Telescope sensitivity threshold, for the rest of the time.
A multifrequency campaign of 3C~111, ongoing in the same period, revealed  an increase of  the mm, optical and  X-ray fluxes in 2008 September-November, interpreted by  Chatterjee et al. (2011)  as due to the passage of a superluminal knot through the jet core.  The temporal coincidence of the mm-optical-X-ray outburst with the GeV activity suggests a co-spatiality of the events, allowing, for the first  time, the  localization of  the $\gamma$-ray dissipative zone in a  FRII jet. 
We argue that the GeV photons of  3C~111 are produced in a compact region confined within 0.1 pc and at a distance of about 0.3 pc from the black hole.
\end{abstract}

\keywords{galaxies: active Ñ galaxies: jetsÑ gamma-rays: galaxies: individual (3C111)}

\section{Introduction}
The Large Area Telescope (LAT) on board the   {\it Fermi} ~ $\gamma$-ray  satellite \citep{Atwood2009} has recently detected  a handful of Misaligned Active Galactic Nuclei (MAGN),  i.e., Radio Galaxies (RGs) and Steep Spectrum Radio Quasars \citep{Abdoa2010}.
In spite of their small number, these $\gamma$-ray emitters are extremely appealing sources, as they  offer a different perspective to approach  the high energy phenomena.

For example, the study of the Spectral Energy Distribution (SED) of  Fanaroff-Riley I (FRI) RGs such as NGC~1275 \citep{Abdo2009a},  M87 \citep{Abdo2009b} and NGC~6251 \citep{migliori} has clearly evidenced the  difficulty of the one-zone homogeneous Synchrotron Self-Compton model (SSC) in describing the jet emission within the AGN  unified scheme.
The SED modeling required low Lorentz  factors ($\Gamma \lesssim  3$), implying  bulk velocities
of the RG jets  much lower than typical values found in BL Lac objects (BLs). This is in contrast with the idea that blazars and MAGN are the same kind of source, only characterized by a different inclination angle of the jet (Urry $\&$ Padovani 1995). As suggested by several authors \citep{Georganopoulos2003, ghi05,  Giannios}, this difficulty can be resolved if the hypothesis of a one-zone homogeneous emitting region is relaxed and a structured jet is assumed.  In these models, an efficient (radiative) feedback between different zones at different velocities can reconcile the apparent FRI-BL discrepancy. 
In the MAGN sample  detected by the {\it Fermi}-LAT, there is a clear predominance of nearby RGs with FRI morphology. High power radio sources (i.e., FRIIs)  with GeV emission  are rare. 3C~111 is indeed the only FRII Radio Galaxy of the sample with a secure $\gamma$-ray  association. 
The rarity of  $\gamma$-ray counterparts of FRIIs could be due to intrinsic jet differences \citep{g2011}.

The  exact localization of the $\gamma$-ray emitting region is however still an open problem.
Fast GeV flux variations, observed on timescales of hours to days (Abdo et al 2010c) have unambiguously attested to the compact nature (but not the localization) of the 
$\gamma$-ray source  in blazars.  Very recently, Agudo et al. (2011a,b) claimed that multiwavelength outbursts, observed in two  BLs, i.e.,   OJ~287 and  AO~$0235+164$, can occur in different zones along the jet, in the stationary core as well as  in knots, even at larger distances ($>14$ pc) from the black hole.  Furthermore, in the Flat Spectrum Radio Quasar  PKS~1510$-$089, the complexity of the radio to $\gamma$-ray variability was interpreted as being due to different events occurring along the jet as a knot propagates \citep{ma10}.   Constraining  the black hole--$\gamma$-ray source distance is a critical issue. Indeed it has a strong impact on the physical models invoked to explain the high energy emission (Tavecchio et al.  2011). 

In Radio Galaxies, the situation is even more complicated. Time variability is difficult to detect at GeV energies, because of poor statistics.  In addition, the discovery of  $\gamma$-ray emission from the radio lobes of Centaurus A shows that GeV photons can also originate from extended structures \citep{abdo10s, yang}.
In only one  case,  the FRI RG NGC~1275,   flux variations on time scales of months  were clearly detected  in the  LAT light curves \citep{Abdoa2010, Kataoka2010}   allowing the compact nature (but not the localization) of the $\gamma$-ray emitting region to be asserted.  
Only in the case of the FRI Radio Galaxy M87,  some clues on the  location were provided by the very high energy (VHE) photons, detected by the ground based Cherenkov telescopes (Aharonian et al. 2003, 2006; Acciari et al.  2008, Albert et al. 2008).
 The TeV flare in 2005 was simultaneous to a strong X-ray burst  associated with the innermost knot in the jet  (Harris et al. 2006).  In 2008, the coincidence of the VHE and radio flares, monitored with the Very Long Baseline Array (VLBA),  confined  the VHE photon production within ten Schwarzschild radii (Acciari et al. 2009). A subsequent TeV flare in 2010 was related to a  flux  increase of the core in X-rays (Raue et al. 2011). The difficulty in the data interpretation does not allow an unequivocal  spatial identification of the emitting region.

In this paper we present the first localization of a  $\gamma$-ray flare in a FRII Radio Galaxy.  The source, 3C~111, was observed by {\it Fermi}  in a high state in late 2008, while a radio-optical-X-ray campaign was ongoing (Chatterjee et al. 2011, hereafter CH11). The  mm-to-$\gamma$-ray  light curves combined with high resolution VLBA images indicate that the GeV flare occurred during the  ejection of a bright plasma knot from the radio core.

\section{The 3C~111 $\gamma$-ray flare}
\subsection{The source}

3C~111 (z=0.0491; Eracleous et al. 2004), optically classified as a Broad Line Radio Galaxy (BLRG) shows  a typical FRII morphology. It exhibits radio lobes, a strong core and  a prominent 120$\arcsec$ one-sided jet,  terminating in the hot spot of the northeast lobe \citep{LP84}.
 At a distance of  196 Mpc,  the linear extension of  the jet is ~ 114~kpc  (1$\arcsec$=948 pc) in projection, assuming $H_0=71$ km s$^{-1}$ Mpc$^{-1}$,  $\Omega_m = 0.27$, and $\Omega_\Lambda = 0.73$.\\  
VLBA observations reveal a one-sided core-jet structure characterized by an  intrinsic half opening angle $\phi \sim 3^{\circ}$ and an inclination angle $\theta \sim 18^{\circ}$ \citep{jor05}. The milliarsecond (mas)  jet contains multiple superluminal components with apparent velocities $\beta_{app}=2-6$ \citep{kad08}, ejected approximately  every 5 months from the core \citep{jor05}.
The accretion disk of this source is very efficient in converting gravitational power into radiation, as attested by the presence of an iron K$\alpha$ line detected by several X-ray satellites (CH11, Ballo et al. 2011).

3C~111 was one of the few sources in the MAGN sample proposed as a possible $\gamma$-ray counterpart of an EGRET detection \citep{hartman08} and subsequently confirmed by  {\it Fermi}. In the first LAT catalog of AGN \citep{Abdob2010},  based on  11 months of data,  the source was associated  with the low latitude   $\gamma$-ray source 1FGL~J0419.0$+$3811 (detection significance $\sigma=4.3$) with a high probability ($P=87 \%$).   3C~111 was still detected with a similar significance after 15 and 24 months of survey \citep{Abdoa2010, kat11}. It is not listed in the  second LAT AGN catalog \citep{ackermann}, where only sources with a detection $\ge 5 \sigma$ are included. 
Indication of possible time variability was first suggested  by Hartman et al. (2008). They noted that the source only occasionally became bright ($F_{100} > 10^{-7}$ phot cm$^{-2}$ s$^{-1}$) and detectable by EGRET, suggesting a low duty cycle for the $\gamma$-ray emission.
The {\it Fermi} light curve, based on 15 months of data, revealed a similar behavior.   3C~111 was detected with a significance larger than 3$\sigma$ only in 2008 August-November \citep{Abdoa2010}.

\subsection{The 24-month $\gamma$-ray light curve}
Taking advantage of the recent improvement of the instrument response functions (IRFs) and the background models,  we produced  a new 0.1-100 GeV light curve of 3C~111 from 2008 August  to 2010 August  using a bin time  of 2 months.
The {\it Fermi}-LAT ScienceTools software (version v9r23p1) was used with the P7SOURCE$\_$V6 set of instrument response functions. The time intervals  when the rocking angle of the LAT was greater than $52^{\circ}$ were rejected and  a cut on the zenith angle of $\gamma$-rays of 100$^{\circ}$  was applied. 
The model for which we calculated a binned likelihood is the combination of point-like and diffuse sources  located within a region of interest (RoI) having a radius of 15$^{\circ}$ and centered on 3C~111.  
 For each point-like source the spectral  slope was frozen to the best fit value provided by the 
{\it Fermi}-LAT Second Source Catalog  \citep{Abdo11}. The RoI model also includes sources falling between 15$^{\circ}$ and 20$^{\circ}$ from the target source, which can contribute to the total counts observed in the RoI due to the energy-dependent size of the point-spread function of the instrument. For these additional sources, normalizations were also fixed to the values of the 2FGL catalog. The spectral model used for 3C~111 is a power law (F=kE$^{-\Gamma}$)
with the spectral slope fixed to the value reported  in 1LAC paper $\Gamma=2.6\pm 0.2$. Note that the $\Gamma$ uncertainties have very negligible effect on the flux calculation.
A Galactic emission model (gal$\_$2yearp7v6$\_$v0.fits) and an extragalactic and instrumental background (isotrop$\_$2year$\_$P76$\_$source$\_$v0.txt) were used to model the background. 
 We evaluated the significance of the detection given by the Test Statistic TS=$2\Delta$log(likelihood) between the models with and without the source (Mattox et al. 1996). When the TS was less than 10, the bin flux value at F $> 0.1$ GeV was replaced by a 2$\sigma$ upper limit, calculated  by  finding the point at which 2$\Delta$log(likelihood)=4 when increasing the  flux from the maximum-likelihood value.

The 24-month light curve confirms the high state of 3C~111 in late 2008 (Fig. 1-{\it left~ panel}): the source appears to spend not more than two months flaring and then drops  below the LAT  sensitivity threshold. 
In order to better constrain the $\gamma$-ray  flare shape,  a more dense LAT time sampling from 2008 August to December was then performed. 
A light curve was generated by dividing the  time interval into bins of 1 month duration and repeating the likelihood analysis for each interval (Fig. 1- {\it right panel}). 
The source is detected with a significance of $\sim 3 \sigma$ (TS=9.7)  only on one occasion, indicating  that the GeV flare peaked between October 4 and November 4.  It is probable that the major $\gamma$-ray activity was at the beginning  of October. 
A likelihood analysis performed using a sampling of  only 15 days shows that  the largest Test Statistic value (TS=7) is observed between October 4 and 19.

\subsection{Localization of the $\gamma$-ray dissipation region}

As a  multiwavelength campaign of 3C~111 (CH11)  was ongoing in the same period monitored by LAT,  it was  possible to compare the $\gamma$-ray data with 
mm-optical-X-ray light curves (Figure 2).  The simultaneity of the flare is impressive: the source luminosity was increasing from mm to X-ray frequencies  exactly when the greatest flux of $\gamma$-ray photons occurred. This is a  clear  indication of  the co-spatiality of the event. 
As attested by the TeV-radio campaigns organized for the RG M87 (Acciari et al. 2009), a fruitful approach to identify the  zone where high energy photons  are produced is to follow the jet dynamics at mas scales. 
We therefore investigated  the VLBA images at 43 GHz made available by the blazar research group at Boston University\footnote{\scriptsize http://www.bu.edu/blazars/} and discussed in CH11.
Looking at Figure 6 of CH11, it appears evident that a deep modification of the core-jet system  took place in late 2008. As deduced by the same authors,  on  October 29 (with an uncertainty of $\sim 26$ days) a new component was ejected from the stationary core. 
In order to quantify the jet dynamics, we analyzed the VLBA images at four epochs, before (2008 September) and after  (2008 November--December,  2009 January)  the appearance of the new  blob. To minimize the free parameters, the model fitting of the ({\it u,v}) data were performed with DIFMAP (Shepherd 1997) using the simplest models (i.e., delta, circular Gaussian, elliptical Gaussian). The model components were added to the residual images with a restored beam size of 0.1 mas (Figure 3). 
This is the angular resolution in the direction of the jet propagation, obtained with a suitable weighting of the visibilities. 
The brightest component  was identified as the stationary core, while the westernmost component, at a distance of $\sim$ 0.1~mas from the core, was associated with a possible counter-jet. The new knot  was emerging in November and appears clearly separated from the core in the successive epochs.
It is beyond the aim of this paper to attest the real nature of the westernmost feature.
However, we note that  our analysis of 8 epochs, from 2008 June to 2009 February, indicates that the candidate counter-jet does not show any significant systematic displacement.
In the assumption that the counter-jet is stationary with respect to the core, we can estimate the position errors from its position scatter. Using the mentioned 8 epochs,  the scatter is 0.005 mas.
The flux densities of the different jet components, when close to each other,  are difficult to disentangle even with the model fitting procedure used here.  For this reason, we preferred to follow the core evolution, considering the variation of the total flux density, including the contribution of the stationary feature, the counter-jet and, when separated,  the new ejected knot.  In 2009 January, a small component between the head of the ejected  blob and the core is evident. Its flux was also included, given the difficulty 
in establishing its nature. It could be a new emerging knot or simply a disruption/elongation of the component ejected in 2008 (see also Gro{\ss}berger et al. 2011). 
The speed of the knot, considering the blob-core distance at three different epochs (2008 November - 2009 January),  is  1.07$\pm0.05$ mas yr$^{-1}$ ($\beta_{app}=3.3$c) and the ejection time, obtained by extrapolating the linear (core-knot distance vs time) fit,  is  $T_0=$2008.80$\pm0.01$. Our $T_0$ is in agreement, within the uncertainties,  with the value of CH11, while our $\beta_{app}$ is smaller, because we considered only the 3 epochs closest to the ejection.  
It is interesting to notice that, assuming an inclination angle of 18$^{\circ}$ and apparent velocity $\beta_{app}=3.3$c,  the arm length ratio\footnote{\scriptsize The arm lenght ratio is the ratio between the  apparent length  of the approaching ($d_1$) and receding ($d_2$) jets, defined as Q=$\frac{d_1}{d_2}=\frac{1+\beta cos(\theta)}{1-\beta cos(\theta)}$, being $\beta$ the bulk jet velocity in units of the speed of light (Longair \& Riley 1979)} is about 20.  As a consequence,  the observed approaching components extending $\sim2$ mas from the core (Figure 3) should be compressed to $\sim0.1$ mas on the receding side. This latter value is consistent  with the unresolved size of the counter-jet in the VLBA images. 

The light curve of the core is shown in Figure 4. The brightness increases from August to November,  following the evolution of the multiwavelength flare (Figure 2), then decreases when the new radio component becomes visible (Figure 3). 

Finally, we note that two other knots could have been ejected by the core at the beginning of 2009 (CH11). These minor events caused no significant amplification of the GeV radiation. It seems that only exceptional jet perturbations can boost the gamma-ray flux above the LAT sensitivity threshold.

\section{Conclusions}

Marscher et al. (2002) in an enlightening paper on the RG 3C~120 noted that a dip in the X-ray light curve is  generally associated with  an expulsion of a bright superluminal radio knot.  It was suggested  that, as in microquasars,  the inner regions of an efficient accretion disk could be accelerated, causing a shock front streaming along the jet.

3C~111 is quite similar to 3C~120. They are both BLRGs probably hosting a disk-corona system (Haardt $\&$ Maraschi 1991).
CH11, exploring the radio optical and X-ray light curves of 3C~111 noted, also for  this source, a strong suppression of X-rays  followed by  the appearance of a new component in the high resolution radio maps. They interpreted the flux increase, quasi-simultaneously observed at mm-optical and X-ray frequencies after the dip, as  the  passage of a new knot through the jet stationary core (see also  Bell $\&$ Comeau 2011 for a  different interpretation).
\noindent
Incidentally, we note that the {\it Suzaku} X-ray satellite observed 3C~111 before the flare, during the dip,  on 2008 August 22, and  {\it XMM-Newton} observed it after the blob ejection in 2009 February 15 \citep{ballo2011, tor12}.  As expected, there is a difference in the 2--10 keV flux by about a factor 2 between the observations. 

An absorption blueshifted feature,  associated with the  resonant Fe {\small XXVI} Ly$\alpha$ transition, was revealed by {\it Suzaku} but not by {\it XMM-Newton}. This absorption line could
originate  in a fast disk-wind (Ballo et al. 2011 and references therein).
If this detection will be confirmed, the episodic presence of hard X-ray absorption features could represent a signature of an ongoing disk-jet perturbation. 

Our  analysis of 24-month {\it Fermi} data reveals that the production of GeV photons is related to  these phenomena. 
First of all, the short time interval of  3C~111 detectability ($\Delta t  \sim 30-60$ days) by the LAT denotes the compactness of the emitting region. Causality arguments show that it cannot be larger than  $R< \Delta t \delta c < \rm$ 0.1 pc ,  if a Doppler factor of $\delta\sim 3$ \citep{jor05} is assumed. 
This immediately allows us to exclude the radio lobes as the main  source of $\gamma$-rays. Note that assuming an extreme time variability of 15 d reduces the radial extent to 0.04~pc.
In addition, 3C~111 was detected by the LAT exactly when  the  mm, optical and X-ray fluxes were rising. The temporal coincidence of the mm-optical X-ray and $\gamma$-ray flares unambiguously implies  a co-spatiality of the event.
Since the outburst of photons is directly connected to the  ejection of a new radio knot,  it is natural  to localize  the $\gamma$-ray source in the radio core region, the size of which cannot be larger than 0.3~pc, corresponding to the VLBA angular resolution  of $\sim$ 0.1~mas.  

The distance between the stationary radio structure and the black hole (d$_{bh-c}$) is unknown. However,  a weak VLBA feature  is seen at a distance of $\sim$ 0.1~mas from the core (d$_{c-cj}$), just on the opposite side of the ejected bright knot. If this is the counter-jet,  as we suggest,  the black hole necessarily lies within 0.1~mas (i.e., d$_{bh-c}$$<$d$_{c-cj}$). As a consequence, the distance of the $\gamma$--ray dissipation region from the central engine has to be of the same order of magnitude or slightly larger  depending on the position of the $\gamma$-ray source with respect to the counter-jet.
This black hole-$\gamma$-ray source distance ($\sim$0.1~mas corresponding to 0.3~pc),  is surely a conservative upper limit.
Indeed a recent estimate of the separation between the black hole and the base of the jet in the RG M87 (Hada  et al. 2011)  indicates shorter distances between these components: the stationary core seems to be at not more than about twenty Schwarzschild radii from the central engine.

\acknowledgments

The \textit{Fermi} LAT Collaboration acknowledges generous ongoing support
from a number of agencies and institutes that have supported both the
development and the operation of the LAT as well as scientific data analysis.
These include the National Aeronautics and Space Administration and the
Department of Energy in the United States, the Commissariat \`a l'Energie Atomique
and the Centre National de la Recherche Scientifique / Institut National de Physique
Nucl\'eaire et de Physique des Particules in France, the Agenzia Spaziale Italiana
and the Istituto Nazionale di Fisica Nucleare in Italy, the Ministry of Education,
Culture, Sports, Science and Technology (MEXT), High Energy Accelerator Research
Organization (KEK) and Japan Aerospace Exploration Agency (JAXA) in Japan, and
the K.~A.~Wallenberg Foundation, the Swedish Research Council and the
Swedish National Space Board in Sweden.

Additional support for science analysis during the operations phase is gratefully
acknowledged from the Istituto Nazionale di Astrofisica in Italy and the Centre National d'\'Etudes Spatiales in France.

The authors are very grateful to the blazar research group at the Boston University, who have made all the multiwavelength data of the 3C~111 campaign available.
The authors  also thank D. Bastieri , G. Tosti, J. Finke, and M. Kadler  for useful discussions. ET acknowledges the Italian Space Agency for financial support (contract ASI/GLAST I/017/07/0).

{}

\begin{figure}
\plottwo{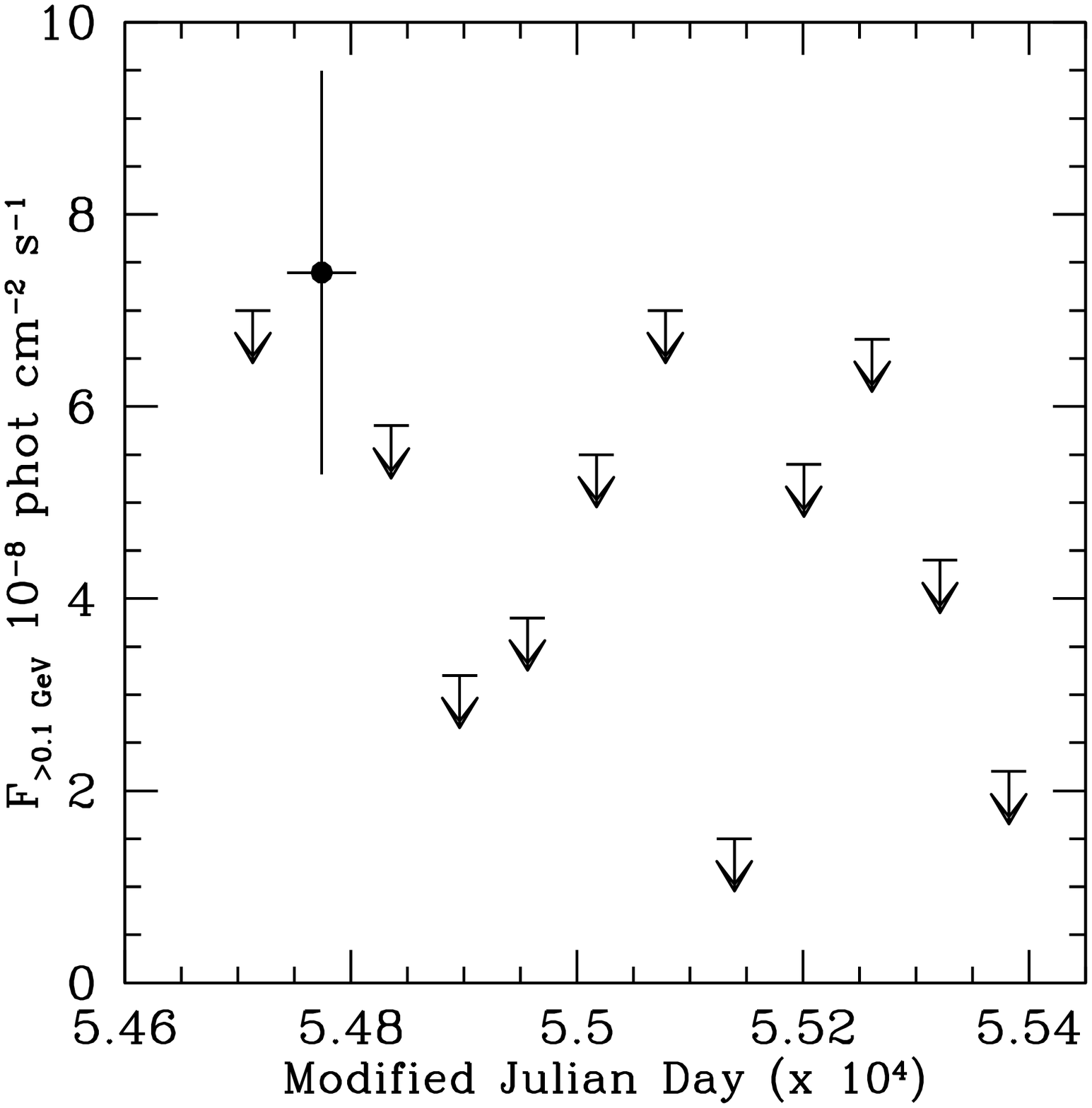}{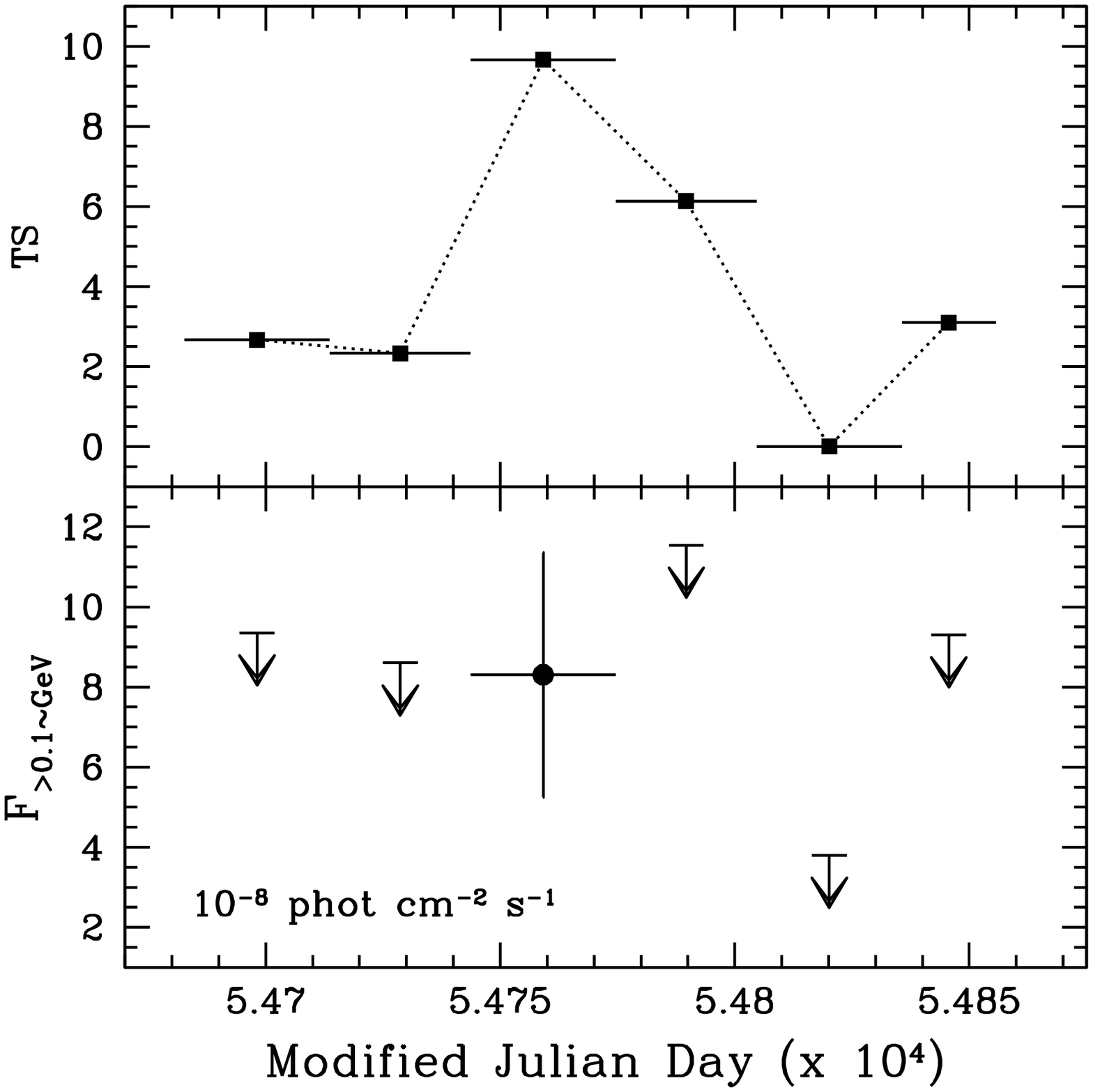}
\caption{{\it Left panel} -- LAT light curve of 3C~111 covering the first 2 years of survey (from 2008 August 4 to 2010 August 4), considering a bin time of 2 months. The source was only detected in 2008 October-November. {\it Right panel} -- One month light curve from 2008 August  to 2009 January (lower box) and  corresponding  Test Statistic for each time interval (upper box). The source is detected with a TS=9.7 ($\sim 3\sigma$) only in 2008 October.}

\end{figure}

\begin{figure}
\plotone{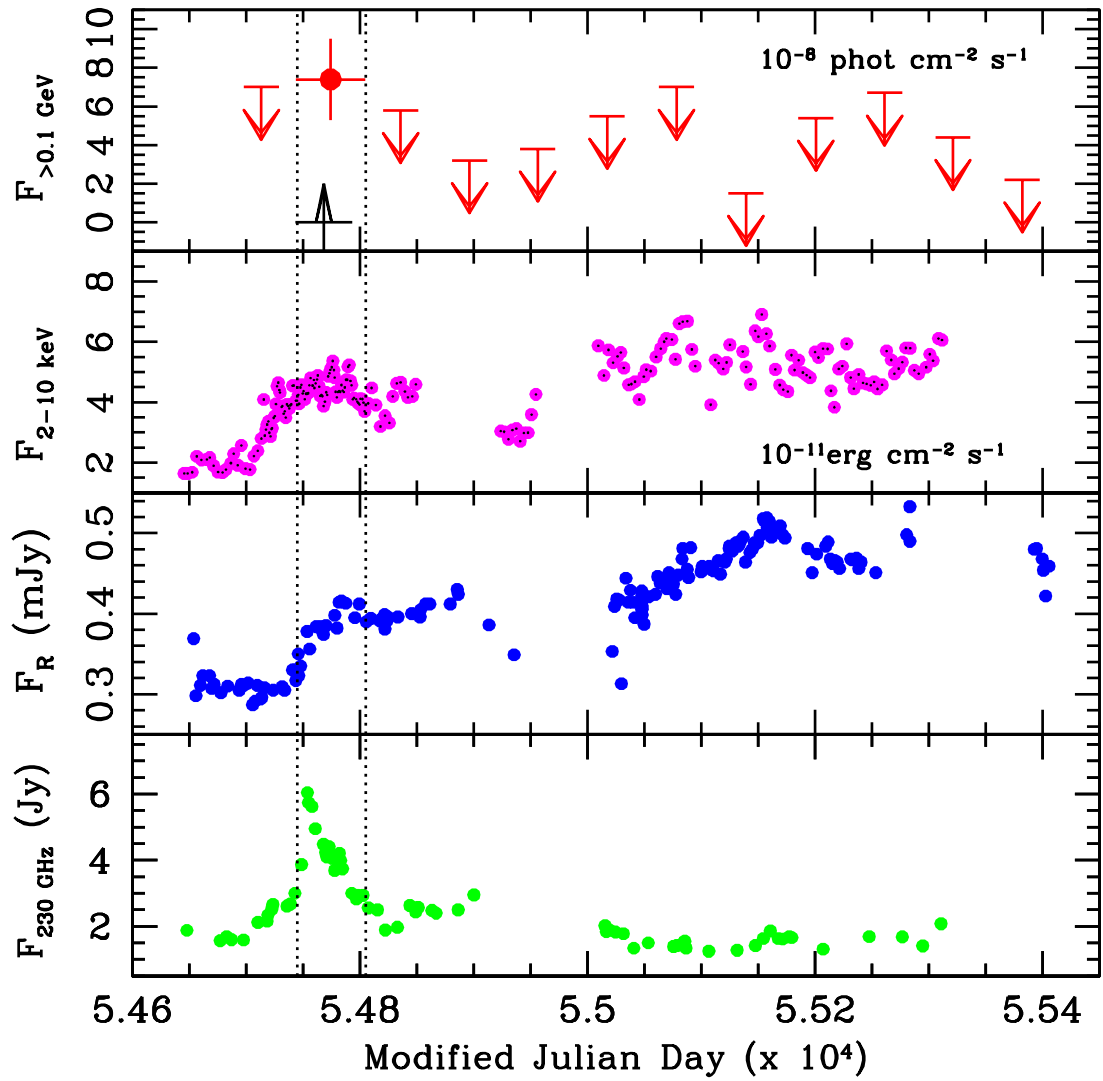}
\caption{Millimeter (230 GHz), optical (R), X-ray (2.4-10 keV) and $\gamma$-ray (0.1-100 GeV) light curves.  The original mm-optical-X-ray data are from CH11. A  detailed study of the correlation among the different light curves is also presented in the same paper.  The black arrow indicates the time when the radio knot was ejected by the core.}
\end{figure}

\begin{figure}
\begin{center}
\includegraphics[width=80mm,height=75mm]{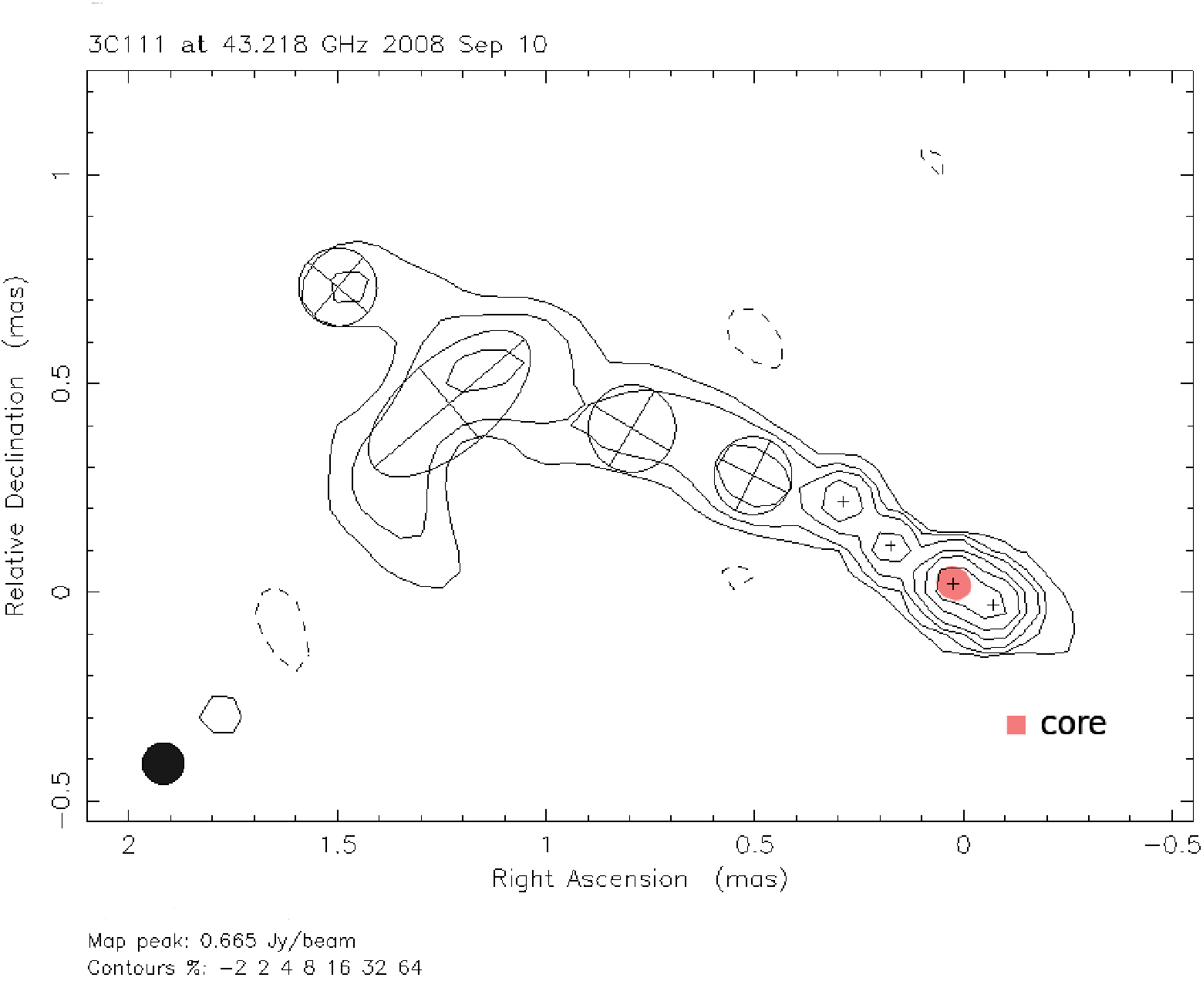}
\includegraphics[width=80mm,height=75mm]{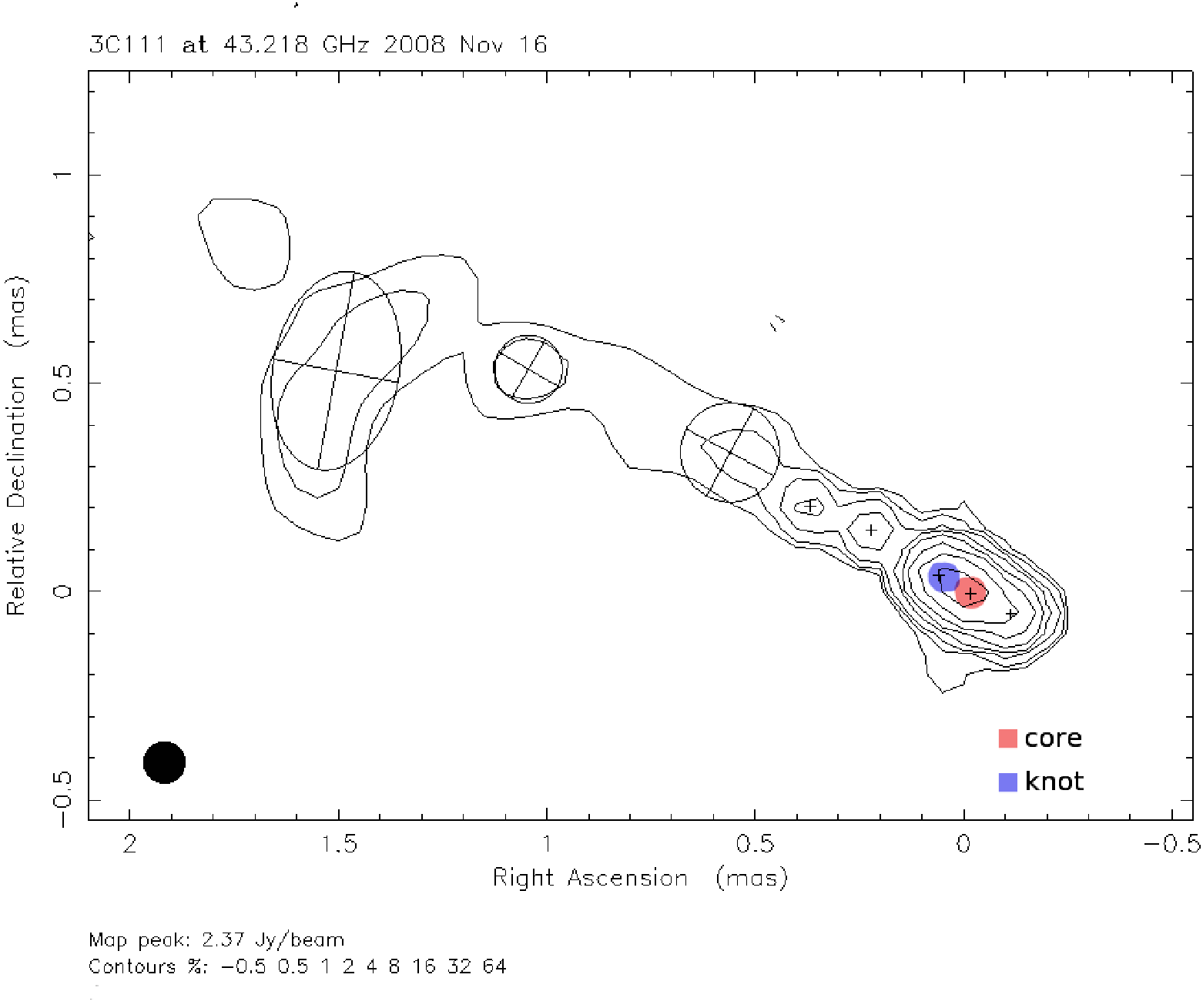}
\includegraphics[width=80mm,height=75mm]{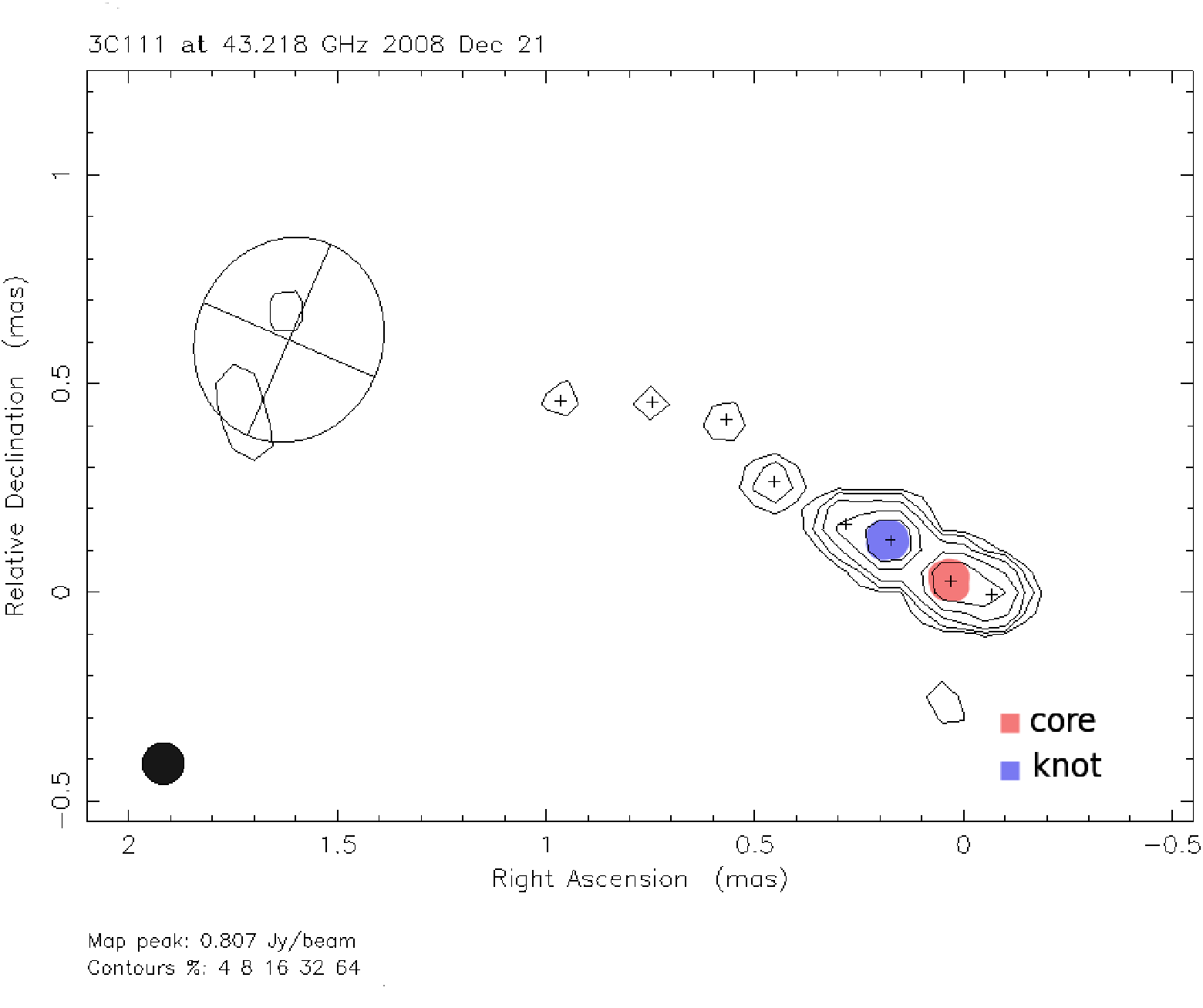}
\includegraphics[width=80mm,height=75mm]{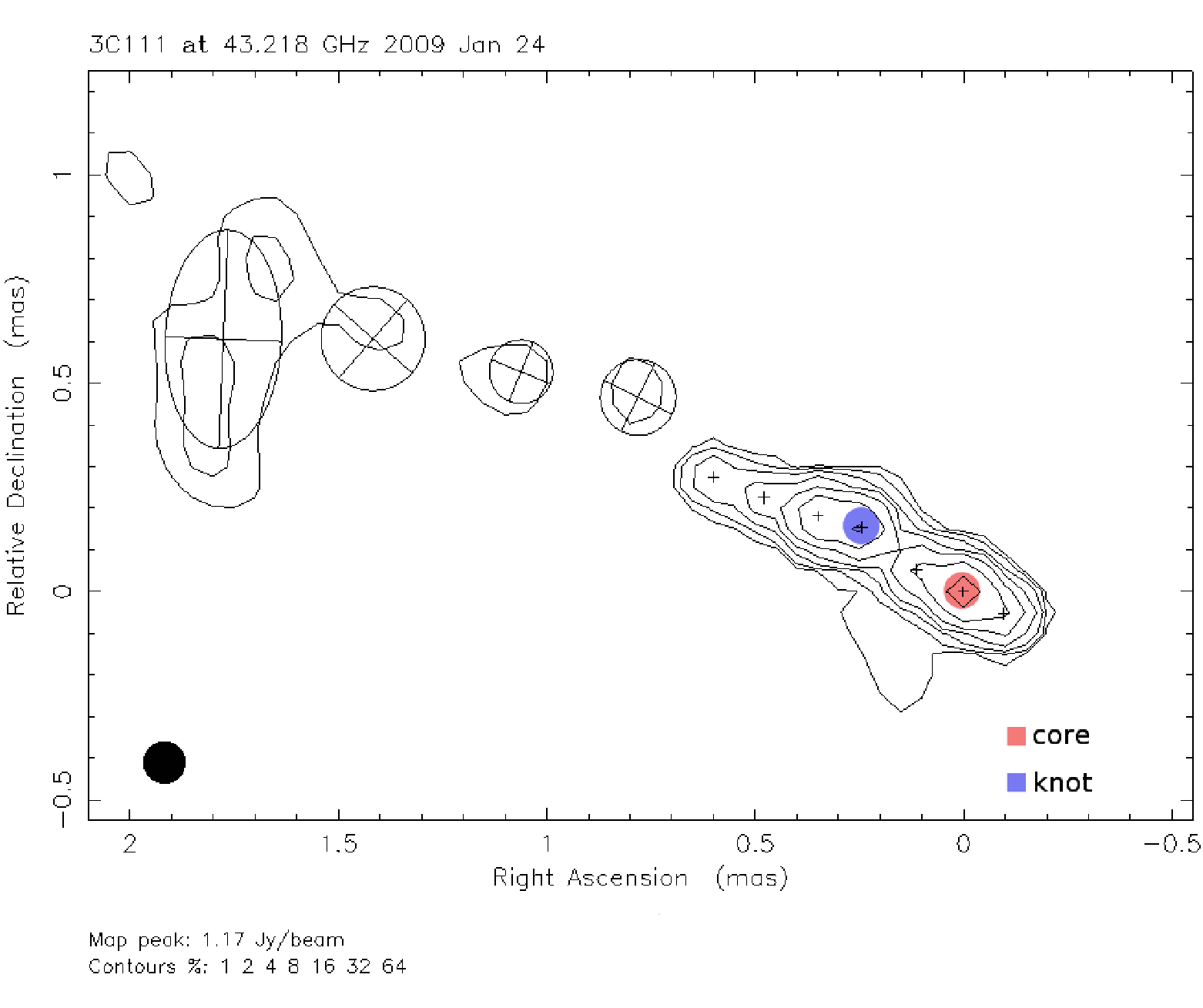}
\caption{VLBA images (43 GHz) at 4 epochs, mapping the emergence of a new knot from the radio core.
The different jet components are shown as crosses (delta profile), circles (circular Guassian profile) and  ellipses (elliptical Guassian profile).  The sequence shows the appearance of a new component (blue filled circle) in 2008 November that is clearly separated  from the core (red filled circle) in 2008 December and 2009 January.}
\end{center}
\end{figure}

\begin{figure}
\plotone{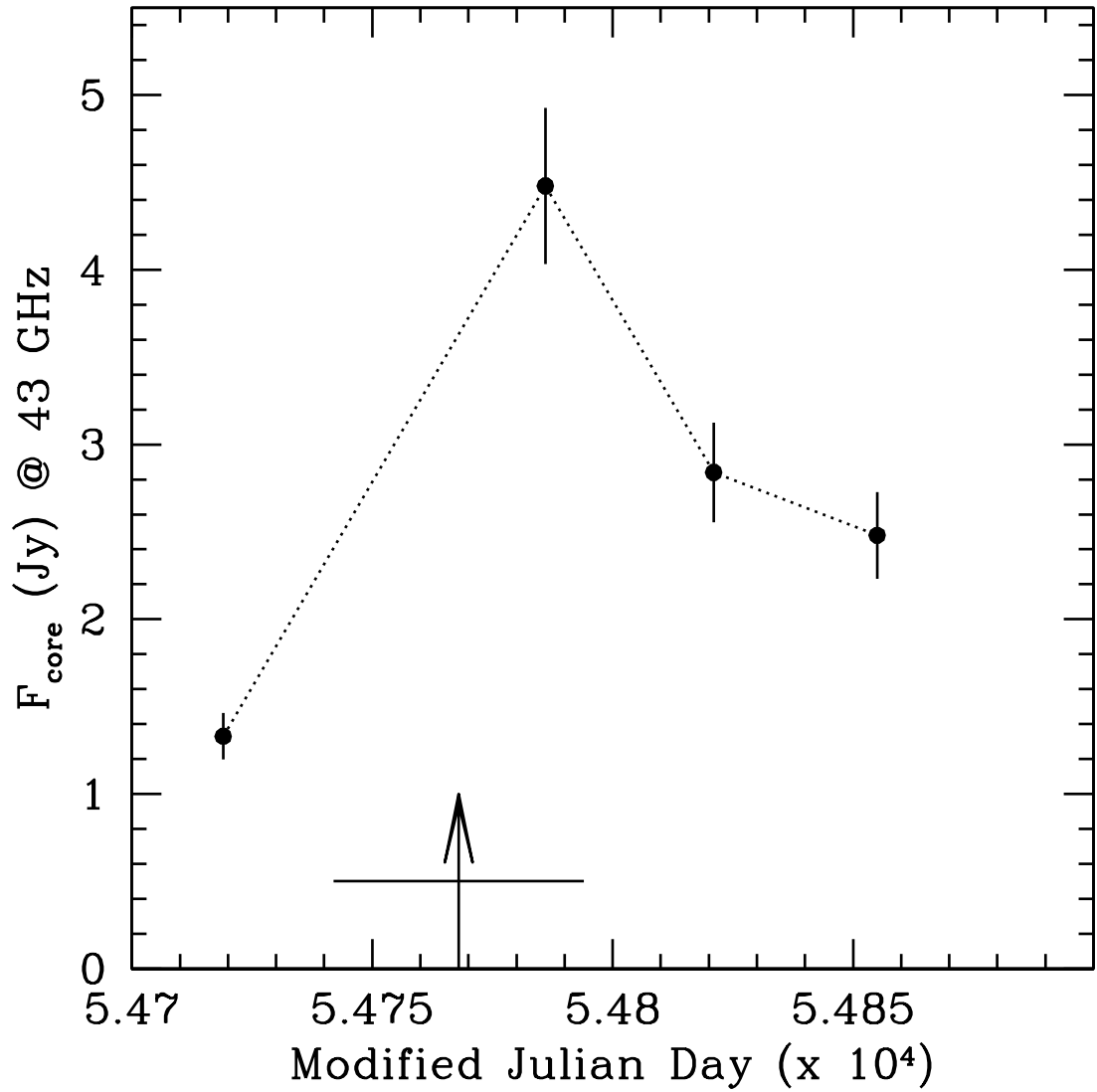}
\caption{Flux Density evolution of the core region as a function of the time from 2008 September 10 to 2009 January 24.
The flux density was estimated in a region including the core, the counter-jet and the emerging knot (see text).  The arrow defines the ejection time as estimated by CH11.}
\end{figure}

\end{document}